\documentclass[12pt,preprint]{aastex}
\usepackage{epsf}
\newcommand{\be}{\begin{displaymath}}
\newcommand{\ee}{\end{displaymath}}
\newcommand{\bea}{\begin{eqnarray}}
\newcommand{\eea}{\end{eqnarray}}

\usepackage{natbib}
\usepackage{hyperref}
\shortauthors{Ma et al.}

\begin{document}

\title{A SUPER-EDDINGTON WIND SCENARIO FOR THE PROGENITORS OF TYPE Ia SUPERNOVAE}

\author{Xin Ma\altaffilmark{1,2,3}, Xuefei Chen\altaffilmark{1,2}, Hai-liang Chen\altaffilmark{1,2,3},
Pavel A. Denissenkov\altaffilmark{4,5}, Zhanwen Han\altaffilmark{1,2}}
\altaffiltext{1}{Yunnan Observatories, Chinese Academy of Sciences, Kunming 650011,
China, maxin@ynao.ac.cn, cxf@ynao.ac.cn}
\altaffiltext{2}{Key Laboratory for the Structure and Evolution of Celestial Objects, Chinese Academy of Sciences,
Kunming 650011, China}
\altaffiltext{3}{University of the Chinese Academy of Science, Beijing 100049, China}
\altaffiltext{4}{Department of Physics \& Astronomy, University of Victoria,
       P.O.~Box 3055, Victoria, B.C., V8W~3P6, Canada}
\altaffiltext{5}{The Joint Institute for Nuclear Astrophysics, Notre Dame, IN 46556, USA}

\begin{abstract}
The accretion of hydrogen-rich material onto carbon-oxygen white
dwarfs (CO WDs) is crucial for understanding type Ia supernova (SN
Ia) from the single-degenerate model, but this process has not
been well understood due to the numerical difficulties in treating
H and He flashes during the accretion. For the CO WD masses from
0.5 to $1.378\,{M}_\odot$ and accretion rates in the range from
$10^{-8}$ to $10^{-5}\,{M}_\odot\,\mbox{yr}^{-1}$, we simulated
the accretion of solar-composition material onto CO WDs using the
state-of-the-art stellar evolution code of {\sc MESA}. For
comparison with the steady-state models (e.g \citet{nskh07}), we
firstly ignored the contribution from nuclear burning to the
luminosity when determining the Eddington accretion rate and
found that the properties of H burning in our accreting CO WD
models are similar to those from the steady-state models, except
that the critical accretion rates at which the WDs turn into red
giants or H-shell flashes occur on their surfaces are slightly
higher than those from the steady-state models. However, the
super-Eddington wind is triggered at much lower accretion rates,
than previously thought, when the contribution of nuclear burning
to the total luminosity is included. This super-Eddington wind
naturally prevents the CO WDs with high accretion rates from
becoming red giants, thus presenting an alternative to the
optically thick wind proposed by \cite{hkn96}. Furthermore, the
super-Eddington wind works in low-metallicity environments, which
may explain SNe Ia observed at high redshifts.
\end{abstract}


\section{Introduction}
\label{sec:intro}

\

Type Ia supernovae (SNe Ia), as successful cosmological
distance indicators, are thought to result from thermonuclear
explosions of carbon-oxygen white dwarfs (CO WDs) in binaries.
However, the nature of their progenitors still remains unclear. It
is very likely that a CO WD grows somehow in mass close to the
Chandrasekhar mass (Ch mass) of $1.378\,{M}_\odot$ (if its
rotation is ignored) and then explodes as a SN Ia (e.g.,
\citealt{nty84}). The growth of a CO WD to the Ch mass limit has
been investigated widely and many progenitor models of SNe Ia have
been proposed in past years (for a detailed review, see
\citealt{wh12,hn00}). The most popular progenitor models are
single-degenerate (SD) and double-degenerate (DD) models. In the
SD model, a CO WD accretes material from its non-degenerate
companion to increase its mass close to the Ch mass (e.g.,
\citealt{wi73,nty84,hp04,wang2009}).  In the DD model, two CO WDs
spiral in due to gravitational wave radiation and eventually
merge, leading to a SN Ia explosion (e.g.,
\citealt{it84,w84,yung94,han98,rui09}). The SD model has become
the favourite one in the past decades because it can explain the
similarities of SNe Ia and can reproduce very well
characteristic observational features of most normal SNe Ia (see a recent review by \citealt{wh12}).

The process of accretion onto the CO WD is crucial for the SD
scenario. At low accretion rates, the accreting WD undergoes
H-shell flashes similar to nova outbursts \citep{gs78,kp94,pk95}.
However, multicycle evolution of the H-shell flash is difficult to
compute, therefore only individual outbursts were followed in most
simulations of novae. If the accretion rate is too high, the
accreted matter will pile up on the surface of the WD and the WD
will evolve into a star like a red giant (e.g.,
\citealt{nomoto1982}; \citealt{nskh07}, hereafter NSKH07). Only in
a narrow regime, is the H-shell burning steady, and the accreted
hydrogen is burnt completely. The SN Ia birth rate based on this
view was much lower than the observationally inferred one (e.g.,
\citealt{yung96,livio00}). To solve this problem, the optically
thick wind regime was proposed to replace the red giant regime
\citep{kh94,hkn96}. In this regime, the H-rich matter is
transformed into He at a critical rate, $\dot{M}_{\rm cr}$, while
the unprocessed material is blown off by the optically thick wind.
It significantly expanded the parameter space for the SN Ia
progenitors, which resulted in a higher theoretical birth rate of
SNe Ia from the SD scenario \citep{hp04,meng09}. However, those
studies did not pay much attention to the super-Eddington wind
which is triggered when the luminosity of the accreting WD exceeds
the Eddington luminosity. The Eddington accretion rate is around
$\sim 10^{-5}\,M_\odot\,\mbox{yr}^{-1}$ (\citealt{nomoto1982};
NSKH07), when only the accretion luminosity was used to evaluate
the radiation pressure on the surface of the WD. \citet{shen07}
proposed that the nuclear burning (H-shell burning during
accretion) near the surface of the WD has also to be considered
when estimating the Eddington accretion rate. They obtained a
range of Eddington accretion rate of
$2-7\times10^{-7}\,M_\odot\,\mbox{yr}^{-1}$ for various WD masses.

In this paper, we use the state-of-the-art stellar evolution code
of {\sc MESA}(version 3635)\citep{pea11,pea13} to simulate the
long-term evolution of CO WDs accreting solar-composition
material. Our aim is to investigate the stability of H-shell
burning on these WDs. The WD masses range from 0.5 to
$1.378\,{M}_\odot$, while the accretion rate is varied from
$10^{-8}$ to $10^{-5}\,M_\odot\,\mbox{yr}^{-1}$. The Eddington
accretion rate is estimated by including the nuclear energy,
gravothermal energy and radiation of the core.

\

\section{Simulation Code and Methods}
\label{sec:method}

\
\

In our study, we
employed the {\sc MESA} {\it default} opacity and EOS tables i.e.
the same as described in Figures 1 and 2 in \citet{pea11}. Our
nuclear network consisted of 21 isotopes, such as $^{1}$H,
$^{3}$He, $^{4}$He, $^{12}$C, $^{13}$C, $^{13}$N, $^{14}$N,
$^{15}$N, $^{14}$O, $^{15}$O, $^{16}$O, $^{17}$O, $^{18}$O,
$^{17}$F, $^{18}$F, $^{19}$F, $^{18}$Ne, $^{19}$Ne, $^{20}$Ne,
$^{22}$Mg, and $^{24}$Mg, coupled by 50 reactions, including those
of the pp chains and CNO cycles. Similar to NSKH07, the He-burning
reactions were neglected for simplicity. Two relevant {\sc MESA}
suite cases were selected for simulations: \texttt{make\_co\_wd}
and \texttt{wd2}.

The suite case \texttt{make\_co\_wd} was used to create CO WD
models. First, we chose a sufficiently massive pre-MS star and
evolved it until the mass of its He-exhausted core reached a value
close to the final WD's mass that we needed, but before its He
shell began to experience thermal pulses. Then, we artificially
removed the envelope, leaving a naked CO core that quickly evolved
to the WD cooling track. We selected hot CO WD models on the top
of the cooling track as initial models for our accretion
simulations\footnote{We encountered some convergence problems when
simulating cold CO WDs, possibly due to the high degeneracy
of materials on the surface of cold WDs and low resolution of WD
models in our study (the total number of mass zones is typically
around 2000). The evolution of cold CO WDs during accretion may
be slightly different from that of the hot models before and at
the onset of H burning, but the difference would become very small
when an equilibrium state is achieved. \cite{iben82} and
\cite{it89} also used hot WD models when they simulated accreting
WDs.}. Using these methods, we have obtained a series of CO WD
models with masses in the range of
$0.5-1.0\,{M}_\odot$. For CO WDs with masses larger
than $1.0\,{M}_\odot$, we increased the mass of our naked
$1.0\,{M}_\odot$ CO WD model by accreting material with its
surface composition until the required mass was reached
\citep{dhbp13}. As a result, we created 12 CO WD models with the
masses of $0.5$, $0.6$, $0.7$,
$0.8$, $0.9$, $1.0$,
$1.1$, $1.2$, $1.25$,
$1.3$, $1.35$ and  $1.378\,{M}_\odot$, with central temperatures of (in units of $10^7{\rm K}$) 7.5, 7.1,
7.0, 6.9, 7.7, 10.0, 10.4, 12.2, 13.7, 16.0, 20.7 and 25.2,
respectively, at the onset of accretion.

The suite case of \texttt{wd2} was used to simulate accretion onto
CO WDs. The accreted material has the solar composition, in
particular, its H and heavy-element mass fractions are $X=0.7$ and
$Z=0.02$. We turned on a {\sc MESA} option that allows to include
the acceleration term in the equation of hydrostatic equilibrium.
We have done a series of simulations for all of our CO WD models
with the accretion rate in the range from
$10^{-8}$ to
$\,10^{-5}\,M_\odot\,\mbox{yr}^{-1}$. Each simulation was carried
out long enough to obtain its detailed accretion properties.

We realize that a real CO WD accretes material from its companion
via an accretion disk. Therefore, the WD should initially
accumulate the accreted material at its equator. However, this
material will probably spread over the entire WD surface on a
dynamical timescale because of a dynamical instability, resulting
in its quasi-spherical distribution \citep{mac83}. This motivates
our assumption of spherically symmetric accretion, which is made
and used for simplicity, like it was done previously in other
similar studies (e.g. \citealt{it89,pk95,hkn96}, NSKH07,
\citealt{shen07}), although multi-dimensional simulations would
provide more consistent results.

If the accretion rate is high enough, the luminosity $L$ of the
accreting WD may exceed the Eddington luminosity \bea {L}_{\rm
Edd}=\frac{4\pi GMc}{\kappa_{\rm T}}, \label{eq:Ledd} \eea where
${M}$ is the WD mass and $\kappa_{T}$ is the Thomson
opacity\footnote{ We use the Thomson opacity here for comparison
with previous work. However, in our calculation, $\kappa_{T}$ was
replaced by the opacity of the photosphere.}, in which case the
super-Eddington wind is triggered.

The total luminosity ${L}$ consists of four parts, namely the
nuclear burning energy, gravothermal energy released by
compression, thermal radiation of the core, and accretion energy
released by the accreted material. The nuclear burning luminosity,
${L}_{\rm nuc}$, is always the main part of ${L}$. It can be
written as
\bea {L}_{\rm nuc}=XQ\dot{M}_{\rm nuc},
\label{eq:nuc}
\eea
where $\dot{M}_{\rm nuc}$, ${X}$ and ${Q}$ are the rate
of the accretion material processed by hydrogen burning, hydrogen
mass fraction in the accreted material, and the energy released
per a unit mass of H transformed into He, respectively.

The accretion luminosity released by accreted material, ${L}_{\rm
acc}$, is
\bea {L}_{\rm acc}=\frac{{GM}\dot{M}_{\rm acc}}{R},
\label{eq:acc}
\eea
where ${M}$ and ${R}$ are WD's mass and
radius, and  $\dot{M}_{\rm acc}$ is  the accretion rate.

By letting $L_{\rm Edd}\,=\,L_{\rm acc}$, the Eddington accretion
rate is $\dot{M}_{\rm Edd}=4\pi{cR}/\kappa_{\rm T}$, which was
used in NSKH07. If $L_{\rm Edd}\,=\,L_{\rm nuc}$ and $\dot{M}_{\rm
acc} =  \dot{M}_{\rm nuc}$ (i.e. the accreted material is burnt
completely), the Eddington accretion rate then becomes
$\dot{M}_{\rm Edd}=4\pi GMc/(\kappa_{\rm T} XQ)$, which was
adopted in \citet{shen07}. Since $L_{\rm nuc}>>L_{\rm acc}$,
the value of $\dot{M}_{\rm Edd}$ obtained in \citet{shen07} was
much lower than that in NSKH07. Recently, \citet{tsyl13} also
obtained a much lower Eddington accretion limit than that of
NSKH07 by setting $L_{\rm Edd}\,=\,L_{\rm nuc}+L_{\rm acc}$.

Usually, the accretion energy is not taken into account in ${L}$
because it is radiated away from the WD's surface very quickly
\citep{tb04}. Here, we define a luminosity ${L}_{\ast}$ as
${L}-{L}_{\rm acc}$ for convenience.

In our study, we first set ${L}_{\rm Edd}\,=\,{L}_{\rm acc}$ as
the wind triggering condition to reproduce the results of NSKH07
and examine the reliability of the method. Then, we let ${L}_{\rm
Edd}\,=\,{L}_{\ast}$ as the wind triggering condition to study the
behaviors of CO WDs during accretion, which is followed by a
discussion of the inclusion of ${L}_{\rm acc}$ to ${L}$.

\

\section{Results}
\label{sec:results}

\subsection{Reproducing Previous Results With the New Method}
\label{sec:rpr}

To directly compare with the results of the steady-state models of
NSKH07, we first employed ${L}_{\rm Edd}\,=\,{L}_{\rm acc}$ as the
triggering condition for the super-Eddington wind. However, the
super-Eddington wind obtained under this condition cannot blow off
a sufficient mass during an H-shell flash and, as a result,
the calculation of the following evolution of the star is very CPU
consumptive. Therefore, all simulations encountering H-shell
flashes were investigated by setting ${L}_{\rm
Edd}\,=\,{L}_{\ast}$, while ${L}_{\rm Edd}\,=\,{L}_{\rm acc}$ was
set for other cases. Figure 1 shows three typical examples of our
simulations: (a) an H-shell flash at a low accretion rate,
$10^{-7}\,{M}_\odot\,\mbox{yr}^{-1}$; (b) steady H burning at an
intermediate accretion rate, $2.1\times
10^{-7}\,{M}_\odot\,\mbox{yr}^{-1}$; and (c) a WD becoming a red
giant at a high accretion rate, $8\times
10^{-7}\,{M}_\odot\,\mbox{yr}^{-1}$ (hereafter, the red-giant
regime). In panels (b) and (c), there are also weak H-shell
flashes when H-rich material is ignited. Such an H-shell flash is
unavoidable at any given accretion rate because H-rich material
accumulated on the WD surface becomes electron-degenerate before
it is initially ignited. Therefore, we first employed ${L}_{\rm
Edd}\,=\,{L}_{\ast}$ as the triggering condition for the
super-Eddington wind at the first H-shell flash, and only after
that we set ${L}_{\rm Edd}\,=\,{L}_{\rm acc}$ to simulate the
following accretion evolution.

Figure 2 shows three boundary curves marked by their corresponding
mass accretion rates: $\dot{M}_{\rm stable}$ separates the steady
H burning from the H-shell flash, $\dot{M}_{\rm RG}$ separates the
steady H burning from the red-giant regime, and $\dot{M}_{\rm
Edd}$ separates the red-giant regime from the super-Eddington
wind. All the boundaries were determined via bisection method for
each WD mass. The results of \cite{it89} and NSKH07 are also
presented in Figure~2 for comparison. It is seen that our results
are very close to theirs. However, some differences exist in the
exact locations of the boundary curves from our and previous
works, which are likely caused by different methods employed.
\cite{it89} used the static envelope analysis, NSKH07 used the
linear stability analysis to investigate the stability of the
steady-state models, while we carried out detailed stellar
evolution calculations that included a realistic accretion
process. A detailed comparison of parameters of the WD models in
the steady H burning regimes obtained in our work and in NSKH07 is
made in Table 1. Again, we see that the two models have very
similar properties.

\

\subsection{The Super-Eddington Wind Scenario}
\label{sec:Edd}

Here, we employ ${L}_{\rm Edd}\,=\,{L}_{\ast}$ as a condition for
triggering the super-Eddington wind to calculate the Eddington
accretion rate ($\dot{M}_{\rm Edd}$) for each WD mass. The results
are shown in Figure~3. We find that the values of $\dot{M}_{\rm
Edd}$ in this case are much lower than those from the previous
works, even lower than the values of $\dot{M}_{\rm RG}$ obtained
in section 3.1. The entire red-giant regime is now replaced by a
new regime that we call the ``super-Eddington wind regime''. In
this new regime, material accreted onto the surface of the WD
first undergoes an H-shell flash. After that, the H burning
becomes steady and then the super-Eddington wind is triggered. The
super-Eddington wind continues to blow off the surface material,
which prevents the envelope from expanding, then the WD will
never become a red giant and its luminosity remains constant at a
value of the Eddington luminosity.

In the super-Eddington wind regime, H burning is stable, and the
accreted material is partially accumulated on the WD surface.
Furthermore, the WD mass growth rate equals approximately to
$\dot{M}_{\rm Edd}$. The extra mass is blown away by the
super-Eddington wind. Thus, the super-Eddington wind is an
alternative to the optically thick wind in preventing an accreting
WD from expanding to a red giant at relatively high accretion
rates. For a convenient use, we have fitted our $\dot{M}_{\rm
Edd}$ and $\dot{M}_{\rm stable}$ data by the following
polynomials: \bea \dot{M}_{\rm Edd}=5.975\times
10^{-6}\left({M}_{\rm WD}^4-3.496{M}_{\rm WD}^3+4.373{M}_{\rm
WD}^2  -2.226{M}_{\rm WD}+0.406\right), \label{eq:MdotEdd} \eea
\bea \dot{M}_{\rm stable}=3.057\times 10^{-7}\left({M}_{\rm
WD}^2-0.386{M}_{\rm WD}+0.027\right), \label{eq:MdotStable} \eea
where ${M}_{\rm WD}$ is in units of ${M}_\odot$, and $\dot{M}_{\rm
Edd}$ and $\dot{M}_{\rm stable}$ are both in units of
$M_\odot\,\mbox{yr}^{-1}$.

In the above analysis, we neglected a contribution of $L_{\rm
acc}$ to $L$. However, for massive WDs, say $1.35{M}_\odot$,
${L}_{\rm acc}\sim 10^3{L}_\odot$ and it contributes to ${L}$ as
much as 10 per cent \citep{fkr85}. The inclusion of ${L}_{\rm
acc}$ in ${L}$ will make the super-Eddington wind to occur more
easily, because an additional source of energy will be
contributing to expelling the accreted material. Given that part
of the accretion luminosity is emitted by the disk, and a part of
the rest of it goes into spinning up of the WD
\citep{lp74,bfgs09}\footnote{The energy deposited to winds is
usually less than $0.5{L}_{\rm acc}$.}, we examined the effect of
${L}_{\rm acc}$ for the following three cases: ${L}_{\rm
Edd}={L}_{\ast}+0.1{L}_{\rm acc}$, ${L}_{\rm
Edd}={L}_{\ast}+0.5{L}_{\rm acc}$, and ${L}_{\rm
Edd}={L}_{\ast}+0.8{L}_{\rm acc}$. The corresponding results are
also shown in Figure 3, in which we see that the Eddington
accretion limit ${does}$ decrease with the inclusion of ${L}_{\rm
acc}$ for massive CO WDs, while little differences exist in
low-mass WDs.

\section{Discussion and Conclusion}
\label{sec:dis}

\

We have investigated the evolution of accreting CO WDs with masses
from 0.5 to $1.378\,M_{\odot}$ for accretion rates from $10^{-8}$
to $10^{-5}\,{M}_\odot\,\mbox{yr}^{-1}$. Our results are
consistent with those from the previous studies of the properties
of H burning on the surfaces of CO WDs during accretion, except
for some differences in the exact locations of the boundaries
between different regimes. We have proposed the super-Eddington
wind regime to replace the optical thick wind regime in preventing
the WD's envelope from expanding at high accretion rates. If a CO
WD accretes material with an appropriate rate (i.e. above
$\dot{M}_{\rm Edd}$), the WD will evolve through the
super-Eddington wind regime. The H in the accreted material is
burnt into He at a rate around $\dot{M}_{\rm Edd}$ and the
unprocessed material is blown away by the super-Eddington wind. If
the underlying He is further burnt into C and O as assumed in the
literature, the WD mass then increases and possibly reaches the Ch
mass. This picture thus provides a potential scenario for the
progenitors of SNe Ia. Note that we assumed a constant accretion
rate in our study, but the strong winds may hit the companion
surface and should affect the mass transfer rate to such an extent
that it could eventually stop \citep{hkn99}.

The characteristics of the super-Eddington wind are similar to
those of the optically thick wind \citep{hkn96}. However, the
efficiency of the optically thick wind strongly depends on the
metallicity because it is driven by a peak in the opacity due to
iron lines, therefore it does not work when the metallicity is
lower than 0.002 \citep{koba98,kobaN09}. On the contrary, the
super-Eddington wind does not significantly depend on the
metallicity. We examined this for two CO WDs (with mass of
1.0 and $1.378{M}_\odot$, respectively) for
${Z}=10^{-6}$, and found that the super-Eddington wind could still
be triggered. The extra mass exceeding the Eddington accretion
rate is blown away by the super-Eddington wind, and the WDs
increase in mass with rates near the Eddington accretion rate (see
Figure 4). This indicates that our super-Eddington wind scenario
may produce SNe Ia at very low metallicities, which could explain
the SNe Ia at high redshifts, e.g. SN UDS10Wil with a redshift of
1.914 \citep{jrr13}, assuming that the metallicity decreases with
the redshift.

Note that the mass-loss rate ($\equiv \dot M_{\rm acc}-\dot
M_{\rm Edd}$ in our study) by super-Eddington wind has an upper
limit, $\dot M_{\rm max}$, above which the super-Eddington wind
cannot blow away all the unprocessed material. From the energy
limit \citep{langer00}, $\dot M_{\rm max}=(\alpha L/L_{\rm
Edd})\,6\times10^{-6}(R/0.01R_\odot)M_\odot{\rm yr^{-1}}$, where
$L$ is the star luminosity and $\alpha$ is the efficiency of
stellar photon luminosity converting into kinetic wind energy
($\alpha=1$ if we assume all the photon energy is used to drive
the wind i.e all photons have been trapped and the WD is
invisible). If $\dot M_{\rm acc} > \dot M_{\rm Edd}+\dot M_{\rm
max}$, the WD may then become a red giant
eventually.\footnote{The mass-loss rate of optically thick
wind \citep{hkn96} is also limited by the energy limit. For
Wolf-Rayet stars, $\alpha \simeq 0.05$ from the study of
\citet{la93}.}

We have not considered abundance mixing induced by H-shell flashes
in our study, since the underlying He shell grows in each H-shell
flash and prevents any dredge-up of heavier nuclei from the core
to the surface zone \citep{kp94}. The abundance mixing might
affect the energy production via nuclear burning, and then the
$\dot{M}_{\rm stable}$ boundary slightly, but not the
super-Eddington wind boundary.

\acknowledgments We acknowledge useful discussions and suggestions
from Xiangdong Li.
This work is partly supported by the NSFC (Nos.11173055, 11033008)
and Yunnan Grant (2012HB037). PAD acknowledges the support of his
research by the NSF grants PHY 11-25915 and AST 11-09174 and by
JINA (NSF grant PHY 08-22648). The computations are made at the
Yunnan Observatories Supercomputing Platform.

%-----------------------------------------------------------------
%\pagestyle{empty}
%\voffset=-3 cm
\begin{figure}
\epsfxsize=10cm
\epsffile[-10 0 450 750]{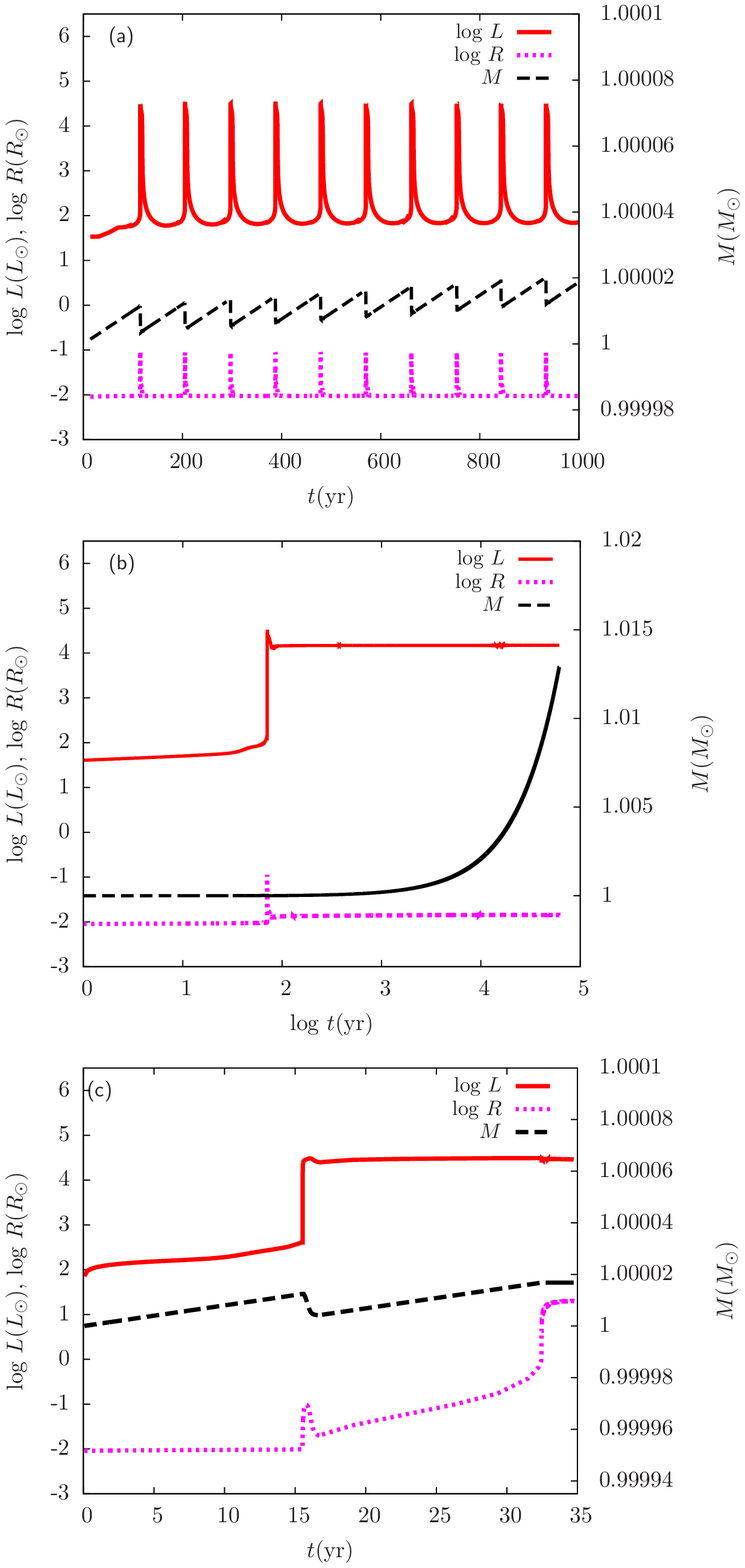}
%\plotone{f1.eps}
\caption{Temporal evolution of the luminosity, radius, and mass of an accreting $1\,{M}_\odot$ CO WD in the H-shell flash
case ($\dot{M}_{\rm acc}\,\sim\,10^{-7}\,{M}_\odot\,\mbox{yr}^{-1}$, panel a), in the steady H burning case
($\dot{M}_{\rm acc}\,\sim\,2.1\times 10^{-7}\,{M}_\odot\,\mbox{yr}^{-1}$, panel b), and in the red-giant regime
($\dot{M}_{\rm acc}\,\sim\,8\times 10^{-7}\,{M}_\odot\,\mbox{yr}^{-1}$, panel c).
         }
\label{fig:f1}
\end{figure}

%-----------------------------------------------------------------

%-----------------------------------------------------------------
%\pagestyle{empty}
%\voffset=-3 cm
\begin{figure}
\epsfxsize=10cm
\epsffile[-10 500 350 750]{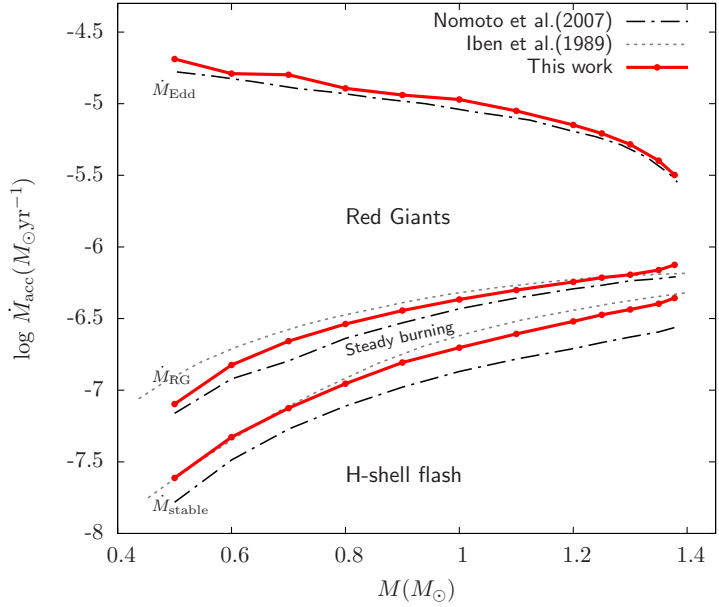}
%\plotone{f2.eps}
\caption{Properties of H burning on the surfaces of accreting CO WDs in the WD mass-accretion rate plane.
The super-Eddington wind is triggered when ${L}_{\rm acc}\,=\,{L}_{\rm Edd}$. The red solid curves are the results of
our calculations. The black dot-dashed curves and the grey dotted curves are from NSKH07 and \cite{it89}, respectively. Note that \cite{it89} did not present their $\dot{M}_{\rm Edd}$ in their paper.
         }
\label{fig:f2}
\end{figure}

%-----------------------------------------------------------------

%-----------------------------------------------------------------
%\pagestyle{empty}
%\voffset=-3 cm
\begin{figure}
\epsfxsize=10cm
\epsffile[-10 480 250 750]{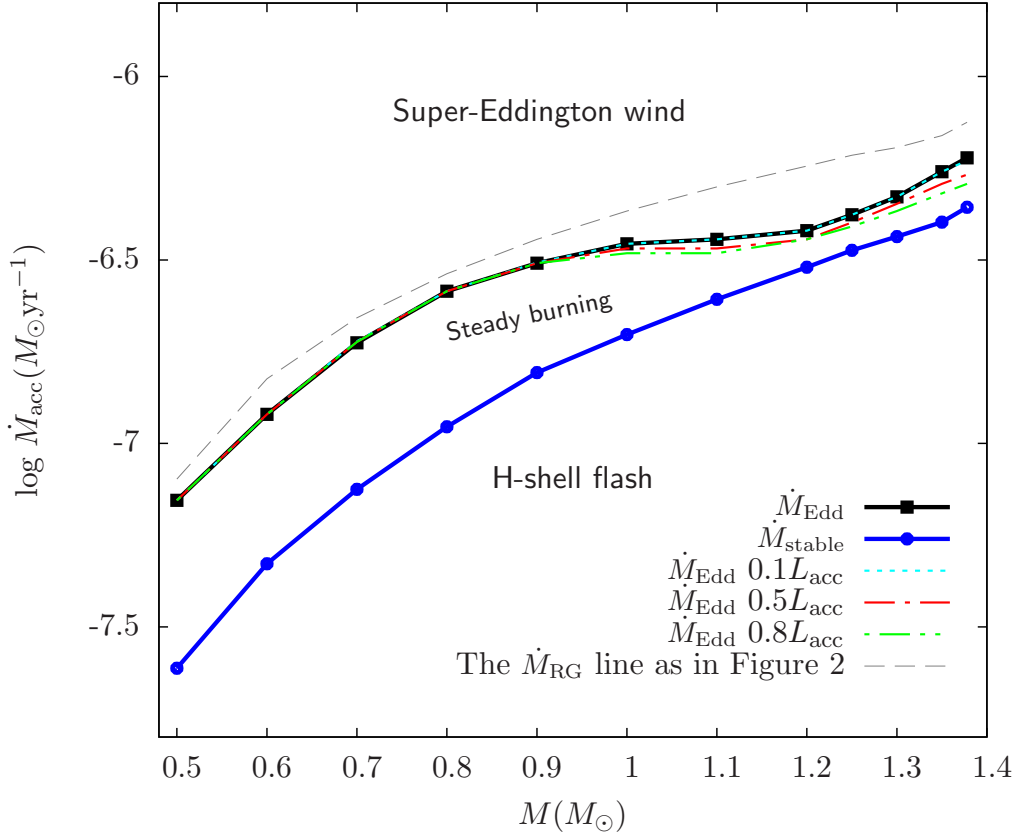}
%\plotone{f3.eps}
\caption{Similar to Figure 2, but the super-Eddington wind is triggered when ${L}_{\ast}\,=\,{L}_{\rm Edd}$ (the thick solid curves).
The light blue dotted, red dot-dashed and green dash-dot-dotted lines show the Eddington accretion limits by adopting ${L}_{\rm Edd}={L}_{\ast}+0.1{L}_{\rm acc}$, ${L}_{\rm Edd}={L}_{\ast}+0.5{L}_{\rm acc}$ and ${L}_{\rm Edd}={L}_{\ast}+0.8{L}_{\rm acc}$, respectively.
The super-Eddington wind starts much earlier and
the boundary of $\dot{M}_{\rm RG}$ does not exist in this condition. For a comparison, the grey dashed curve shows
the boundary of $\dot{M}_{\rm RG}$ presented in Figure 2.
         }
\label{fig:f3}
\end{figure}

%-----------------------------------------------------------------

%-----------------------------------------------------------------
%\pagestyle{empty}
%\voffset=-3 cm
\begin{figure}
\epsfxsize=10cm
\epsffile[-10 530 250 750]{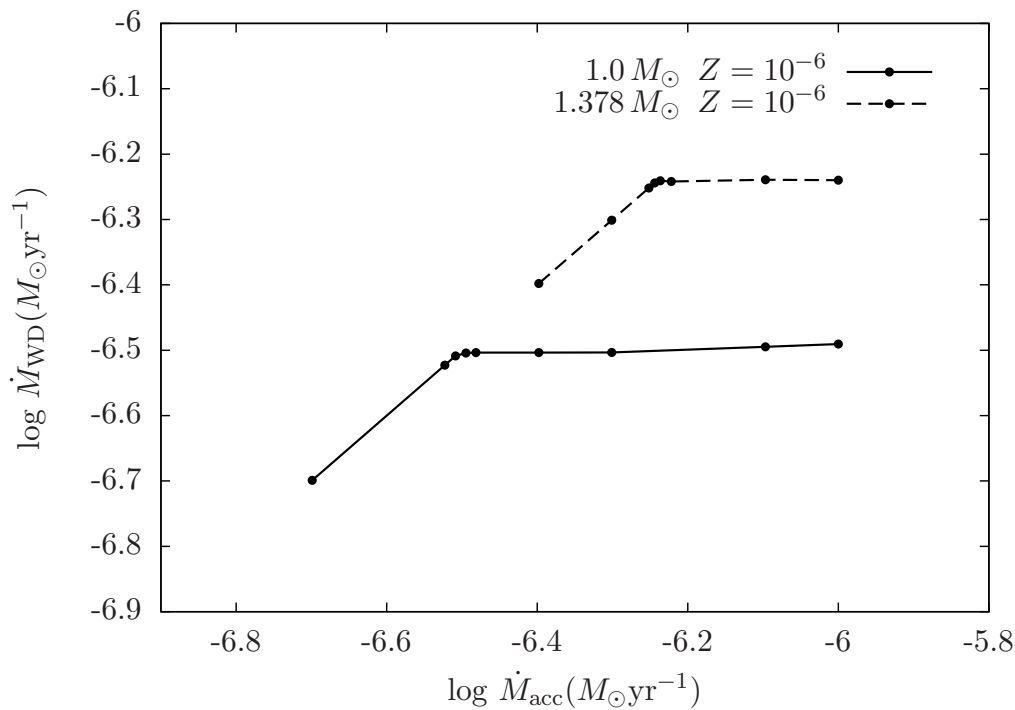}
%\plotone{f4.eps}
\caption{The CO WD mass growth rate, $\dot M_{\rm WD}$, versus
accretion rate at a metallicity $10^{-6}$. The Eddington accretion
rate is $3.1\times 10^{-7}\,{M}_\odot\,\mbox{yr}^{-1}$ for a
$1\,{M}_\odot$ CO WD, and $5.7\times
10^{-7}\,{M}_\odot\,\mbox{yr}^{-1}$ for a $1.378\,{M}_\odot$ CO
WD. If the accretion rate is higher than the Eddington accretion
rate, the extra mass is blown away by the super-Eddington wind and
the mass growth rate of the WD approximately equals to the
Eddington accretion rate.
         }
\label{fig:f4}
\end{figure}

%-----------------------------------------------------------------

%-----------------------------------------------------------------

\clearpage
\begin{deluxetable}{cccccccc}
\tablecaption{Characteristics of two WD models during steady H burning
\label{tab1}
}
\tablewidth{0pt}
\tablehead{
\colhead{Property}&
\colhead{NSKH07} &
\colhead{This work} &
\colhead{NSKH07} &
\colhead{This work}
}
\startdata
${M}_{\rm WD}({M}_\odot)$ & 0.8 & 0.8 & 1.0 & 1.0 \\
$\Delta{M}_{\rm env}({M}_\odot)$ & $2.0\times 10^{-5}$ & $2.0\times 10^{-5}$ & $3.9\times 10^{-6}$ & $6.8\times 10^{-6}$\\
$\dot{M}_{\rm acc}(M_\odot {\rm yr}^{-1})$ & $2.2\times 10^{-7}$ & $2.2\times 10^{-7}$ & $2.5\times 10^{-7}$ & $2.5\times 10^{-7}$ \\
log $r_{\rm H}(R_\odot)$ & -1.888 & -1.905 & -2.034 & -2.029  \\
log $T_{\rm H}$(K)   & 7.89  & 7.85 & 7.93 & 7.88 \\
log $\rho_{\rm H}$(g cm$^{-3}$) & 1.01 & 1.01 & 1.20 & 1.20  \\
log $P_{\rm H}$(dyn cm$^{-2}$)  & 17.31  &  17.17  & 17.50  & 17.34  \\
log $L_{\rm nuc}(L_\odot)$ & 4.188 &  4.177 &  4.243 &  4.245   \\
log $L(L_\odot)$         & 4.218 &  4.185 &  4.274 &  4.256   \\
log $R(R_\odot)$ & -1.255 &  -1.515  & -1.792  & -1.834      \\
log $T_{\rm eff}$(K) & 5.444 &  5.565 &  5.726 &  5.743     \\
\enddata
\tablecomments {${M}_{\rm WD}$ is the WD mass, $\dot{M}_{\rm acc}$ is the accretion rate, and $\Delta{M}_{\rm env}$ is
the mass of the H-rich envelope. $r_{\rm H}$, $T_{\rm H}$, $\rho_{\rm H}$ and $P_{\rm H}$ are the radius, temperature,
density and pressure, respectively, at the bottom of the H-rich envelope. $L$, $L_{\rm nuc}$,
$R$ and $T_{\rm eff}$ are the luminosity, nuclear burning luminosity, radius and effective temperature,
respectively, at the surface.}
\end{deluxetable}

%-----------------------------------------------------------------

%%$\Delta M_{\rm env}(M_\odot)$ & $2.0\times 10^{-5}$ & $2.3\times 10^{-5}$ & $2.5\times 10^{-7}$ & $10^{-5}$ \\%%

\end{document}